\documentclass[preprint,review,11pt]{elsarticle}

\usepackage{graphicx}
\usepackage{amssymb}
\usepackage{amsthm}
\usepackage{lineno}
\usepackage{changes}
\usepackage{color,soul}
\usepackage[export]{adjustbox}   
\usepackage{subfig}
\usepackage{etoolbox}

\usepackage[colorinlistoftodos]{todonotes}
\usepackage{mathtools}
\usepackage{booktabs}
\usepackage{array}
\usepackage{url}
\usepackage[section]{placeins}

\makeatletter
\AtBeginDocument{%
  \expandafter\renewcommand\expandafter\subsection\expandafter{%
    \expandafter\@fb@secFB\subsection
  }%
}
\makeatother

\usepackage{bm}

\definecolor{lightblue}{rgb}{.90,.95,1}
\definecolor{darkgreen}{rgb}{0,.5,0.5}

\definecolor{lightgreen}{rgb}{.90,1,0.90}

\newcolumntype{P}[1]{>{\centering\arraybackslash}m{#1}}

\usepackage{fullpage}

\newtoggle{extra_spacing}
\newtoggle{line_numbers}

\toggletrue{extra_spacing}

\togglefalse{line_numbers}

\iftoggle{extra_spacing}
{
    \newcommand{\stretchval}{1.2}
    \renewcommand{\baselinestretch}{\stretchval}
}{}

\graphicspath{ {./images/} }

\journal{ }

\begin{document}

\begin{frontmatter}
\title{Data-Driven, Physics-Based Feature Extraction from Fluid Flow Fields}
\author{Carlos Michel\'en Str{\"o}fer\corref{corauth}}
\cortext[corauth]{Corresponding author.}
\ead{cmich@vt.edu}
\author{Jinlong Wu\corref{jl}}
\author{Heng Xiao\corref{hx}}
\ead{hengxiao@vt.edu}
\author{Eric Paterson\corref{egp}}
\address{Kevin T. Crofton Department of Aerospace and Ocean Engineering, Virginia Tech, Blacksburg, VA 24060, USA}

\begin{abstract}
    Feature identification is an important task in many fluid dynamics applications and diverse methods have been developed for this purpose. 
    These methods are based on a physical understanding of the underlying behavior of the flow in the vicinity of the feature. 
    Particularly, they rely on definition of suitable criteria (i.e. point-based or neighborhood-based derived properties) and proper selection of thresholds. 
    For instance, among other techniques, vortex identification can be done through computing the Q-criterion or by considering the center of looping streamlines.  
    However, these methods rely on creative visualization of physical idiosyncrasies of specific features and flow regimes, making them non-universal and requiring significant effort to develop. 
    Here we present a physics-based, data-driven method capable of identifying any flow feature it is trained to. 
    We use convolutional neural networks, a machine learning approach developed for image recognition, and adapt it to the problem of identifying flow features. 
    The method was tested using mean flow fields from numerical simulations, where the recirculation region and boundary layer were identified in a two-dimensional flow through a convergent-divergent channel, and the horseshoe vortex was identified in three-dimensional flow over a wing-body junction. 
    The novelty of the method is its ability to identify any type of feature, even distinguish between similar ones, without the need to explicitly define the physics   (i.e. through development of suitable criterion and tunning of threshold). 
    This provides a general method and removes the large burden placed on identifying new features. 
    We expect this method can supplement existing techniques and allow for more automatic and discerning feature detection. 
    The method can be easily extended to time-dependent flows, where it could be particularly impactful. 
    For instance, it could be used in the identification of coherent structures in turbulent flows, a hindrance in the ongoing effort to establish a link between coherent structures and turbulence statistics. 
\end{abstract}


\end{frontmatter}

\iftoggle{line_numbers}
{
    \linenumbers
}{}

\section{Introduction}
    \label{S:intro}

    Feature detection is an important component of data post-processing in fluid dynamics experiments and simulations, and plays an important role in our physical understanding of flow phenomena and fluid-structure interactions. 
    In this study we use convolutional neural networks, a machine-learning approach developed for image recognition, to automatically detect features of interest in fluid flow fields. 
    As with traditional methods, feature detection is done by identification of patterns within the relevant physics-based scalar fields of the flow. 
    Unlike traditional methods, the specific patterns to use are not explicitly specified, but inferred from a set of human-labeled training examples. 

    \subsection{Feature Detection in Fluid Flow Fields}
        A flow feature is a physically meaningful structure within the flow, that is of interest for the application at hand. 
        Examples include recirculation regions, shed vortices, boundary layers, and shock waves.  
        Applications of flow feature extraction include fundamental physical understanding of flow dynamics (e.g. relation between coherent structures and turbulence dynamics\cite{hussain_coherent_1986,robinson_coherent_1991}), engineering design (e.g. reduction of shock wave drag\cite{bushnell_shock_2004}), on-line steering of large simulations (e.g. feature-based adaptive mesh\cite{van_rosendale_floating_1995,kamkar_feature-driven_2011}), among others. 
        Despite the important role flow features play in so many applications, the task of accurately identifying these features remains a laborious one. 
        
        Many methods exist for identifying features based on an understanding of the underlying physics of the phenomena. 
        As an example, methods for identifying vortices include identification based on local field values (e.g. vorticity, helicity, Q-criterion), as well as methods based on global flow properties (e.g. curvature center, looping streamlines)\cite{post_state_2003}. 
        The disadvantages with these methods are that they are specific to the feature in question, limited to certain types of flows (e.g. external flows, turbo-machinery) \cite{post_state_2003}, and rely on creative visualization of the physical phenomena at hand. 
        This has lead to many disjoint methods particularly suited for very specific problems, and places a large burden on developing methods for identifying new features. 

\begin{figure}[htb]
          \centering

          \subfloat[Coherent structures from large eddy simulation, reproduced with permission from Fr\"ohlich et al.~\cite{frohlich_highly_2005}]{
            \includegraphics[width=0.65\textwidth,valign=c]{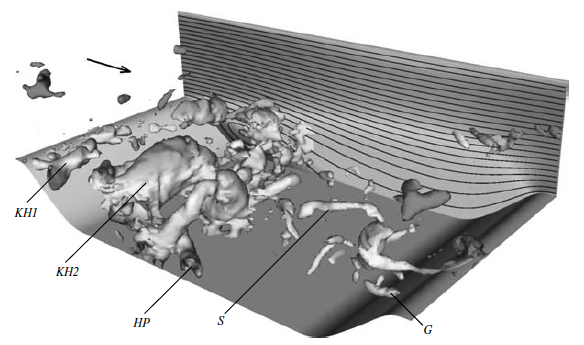}
            \label{fig:structures_3d}
            \vphantom{\includegraphics[width=0.55\textwidth,valign=c]{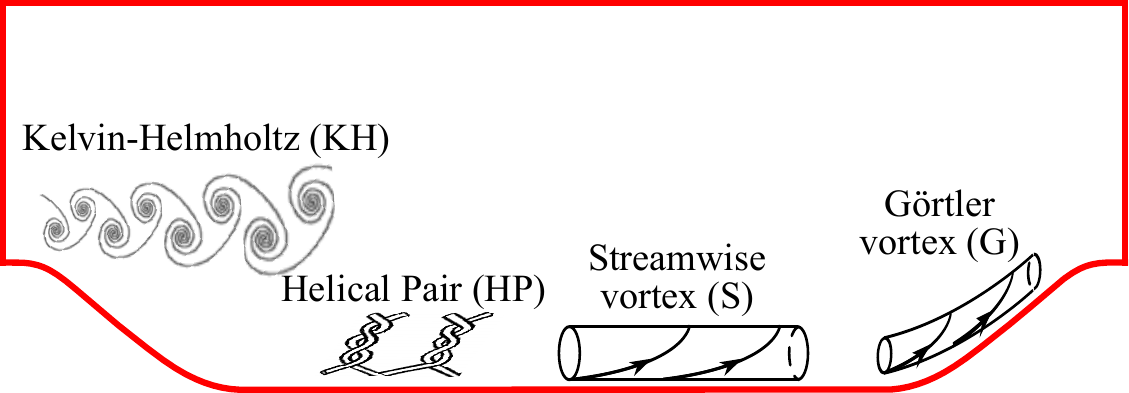}}
         } \quad
         \subfloat[Schematic of coherent structures.]{
            \includegraphics[width=0.6\textwidth,valign=c]{structure-schematic.pdf}
            \label{fig:structures_2d}
         }
        \caption{Distinct vortex structures in the flow over periodic hills as identified by using isosurfaces of pressure fluctuations from large eddy simulation data (a) reproduced with permission from Fr\"ohlich et al.~\cite{frohlich_highly_2005}, and a schematic representation of such coherent structures (b). The types of vortices shown are: Kelvin-Helmholtz vortices (KH), helical pairs of span-wise vortices (HP), streamwise vortices (S), and G\"{o}rtler vortices (G).}
        \label{fig:structures}
        \end{figure}

        As a specific example of the need for improved flow feature extraction we consider the problem of turbulence. 
        Turbulent flows contain temporally-coherent structures (flow features) at various scales. 
        These coherent structures are known to play an important role in mass, momentum, and energy transport \cite{hussain_coherent_1986,robinson_coherent_1991} and subsequently have an effect on overall flow dynamics. 
        Despite this fact, existing turbulence models used for predicting flow separation rely on time-averaged quantities and ignore the presence of coherent structures. 
        For instance, Fr\"ohlich et al.~\cite{frohlich_highly_2005} found that the ``splatting'' of large-scale eddies originating from the shear layer results in a  high-level of spanwise velocity fluctuations in the post-reattachment, which existing turbulence models fail to capture. 
        Figure~\ref{fig:structures} shows an example of these turbulent coherent structures in a flow over periodic hills, showing the large number of distinct types of features and scales present in turbulent flows. 
        For instance, Figure~\ref{fig:structures} shows Kelvin-Helmoltz vortices in the leeward side of the hill and a G\"{o}rtler vortex in the windward side.

        Fundamental understanding of turbulence and development of better predictive models requires establishing a link between organized fluid motion and turbulence statistics. 
        The failure to establish such a link is largely attributed to the lack of reliable data and the lack of effective feature identification techniques for extracting physically significant coherent structures. 
        Effective feature identification would require the ability to automatically search through large time-resolved datasets, search for structures at many scales, and differentiate between similar features (e.g. different types of vortical structures as shown in Figure~\ref{fig:structures}). 
        Existing methods, such as visualization of contours and isosurfaces of flow field variables (e.g, $u'$, $p'$, $\lambda_2$, and $Q$ criterion), require careful choice of criterion and threshold which can be flow and structure dependent.
        This process requires iterative visualization and relies on expert prior knowledge of the specific form of the structure, making it unpractical for large number of time-resolved simulations.
        Moreover, previous studies~\cite{breuer09flow,frohlich_highly_2005} reported that the identification with existing physics-based approaches turned out to be even more difficult at high Reynolds number flows, where large-scale structures are likely to be overwhelmed by small-scale eddies. 
 
        In this paper we provide a universal method for flow feature identification that could replace or supplement the myriad of existing disjoint methods and remove the large burden placed on developing methods for identifying new features.  
        The method is based on convolutional neural networks, a machine learning algorithm used in image recognition, and learns to find structure within the data solely from the examples it is trained with. 
        The approach is fundamentally different to existing methods in that it is data-driven rather than based on explicitly defined physics. 
        This is still physics-based in that the features are identified by searching for relevant structures within scalar fields of physical quantities, but these relevant structures are inferred from the data rather than explicitly defined. 
        This has the advantage of being a general approach, and allowing for the identification of new features for which explicitly defined physics approaches do not exist. 
        The method also has the potential of being more discerning between similar features than existing methods, provided sufficient training data. 
        While flow feature extraction from time-resolved flow fields (e.g. coherent structures in Figure~\ref{fig:structures}) is an important future application, in this initial study we extract flow features from mean flow fields.

    \subsection{Machine Learning for Object Detection in Images}
\begin{figure}[htb]
         \centering
         \subfloat[Object detection.]{
            \includegraphics[width=0.2\textwidth,valign=c]{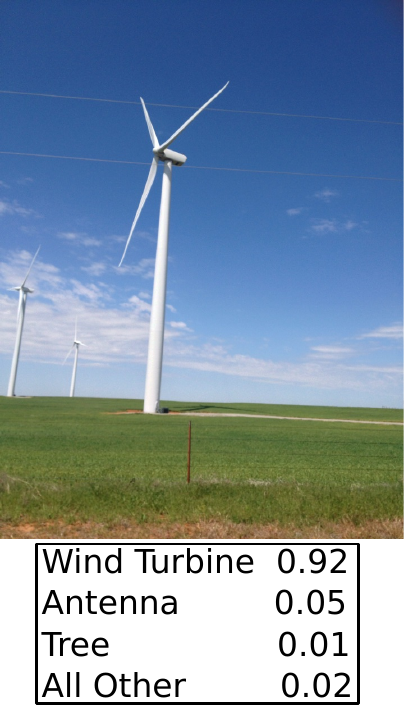}
            \vphantom{\includegraphics[width=0.2\textwidth,valign=c]{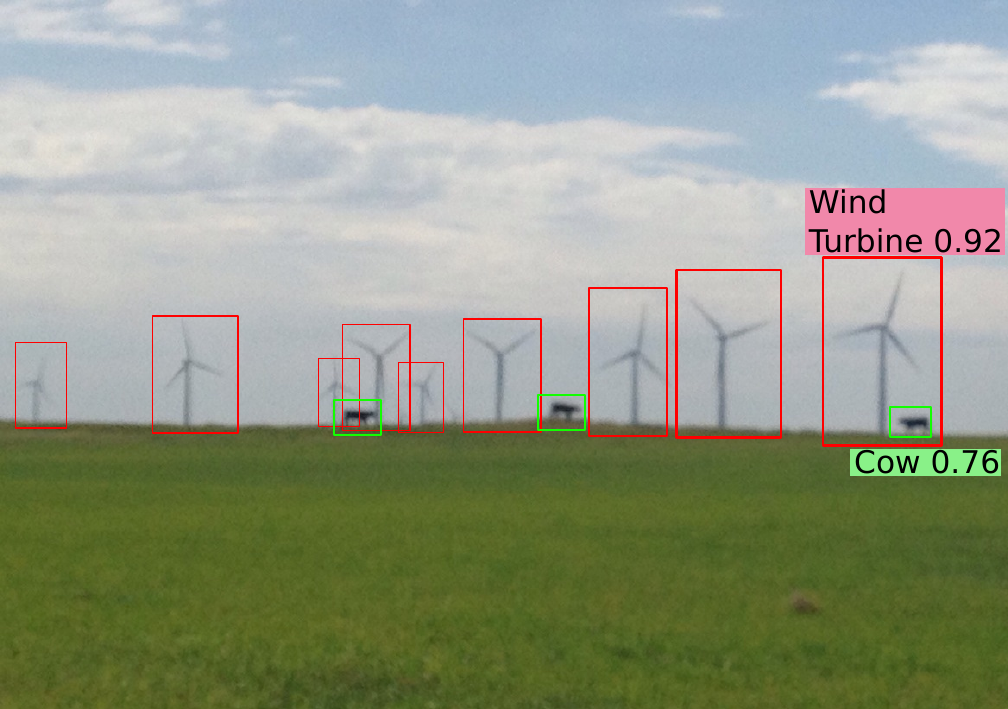}}
            \label{fig:imgrec_det}
         } \quad
         \subfloat[Image recognition.]{
            \includegraphics[width=0.5\textwidth,valign=c]{imagerec_det.pdf}
            \label{fig:imgrec_cla}
         } 
         \caption{Fabricated examples of image recognition (a) and object detection (b) tasks in images. The number indicates the probability of the region or image being an instance of the specified class. The labels for only two of the ``detected'' objects in (b) are shown.}
         \label{fig:imgrec}
        \end{figure}

        Machine learning is a class of algorithms that makes decisions based on learned parameters from data, rather than from explicit instructions. 
        The machine learning algorithms used in this paper were adapted from image recognition and object detection, and an analogy is made between these tasks and flow feature detection. 
        \emph{Image recognition} consists of classifying an image of a single object as belonging to some class (e.g. car, person). 
        \emph{Object detection} consists of finding and classifying all objects within an image that contains more than just a single object. 
        Fabricated examples of image recognition and object detection are shown in Figure~\ref{fig:imgrec}. 
        Both of these tasks can be achieved using different machine learning algorithms. 
        In this paper we consider the object detection R-CNN method developed by Girshick et al.~\cite{girshick_rich_2014, ren_faster_2015} and adapt it to the problem of flow feature extraction. 
        The R-CNN method consists of two main steps, a region proposal step that identifies sub-images containing objects of interest, and an image recognition step using convolutional neural networks (CNN) that classifies those sub-images into appropriate classes. 

        Neural networks are a class of machine learning algorithms that are universal estimators. 
        That is they can be used to estimate any function to any desired accuracy by choice of complexity. 
        Neural networks consist of sequential layers of variables, referred to as neurons, with the first and last layer corresponding to the inputs and outputs respectively. 
        The number of neurons in the first and last layer are determined by the dimensions of the inputs and outputs, while the number of intermediate layers and the size of each are free hyper-parameters chosen to achieve a desired level of complexity.  
        Specific non-linear mappings are defined between adjacent layers, hence an input can be propagated forward through the network resulting in an output. 
        The parameters of these non-linear mappings are learned from a set of training data. 
        The specific choice of layers and mappings is referred to as the network architecture.  
        While Neural networks are universal estimators, estimating an image classification function can require complex architectures with a large number of trainable parameters. 
        
        Convolutional neural networks (CNN) \cite{lecun_gradient-based_1998,simmard_best_2003,nielsen_neural_2015} are a class of neural networks that exploit known functional structures in the image classification problem in order to simplify the network architecture and speed up training. 
        Specifically CNNs use convolutional layers which exploit translational invariance (e.g. whether a region of the image contains an object of interest does not depend on the location of the region relative to the whole image), greatly reducing the number of neurons. 
        Informally a convolutional layer can be thought of as scanning the image searching for different local-features (e.g. horizontal lines, gradients), and creating a local-feature map for each. 
        Convolution layers can be used sequentially, identifying increasingly more complex patterns with the largest scale patterns used for the final classification. 

    \subsection{Machine Learning for Feature Detection in Fluid Flow Fields}
        The problem of flow feature detection is analogous to the problem of object detection in that flow features make a subset of the input and can be found anywhere within the flow. 
        For this reason the authors believe the R-CNN method, and CNNs in general, are well suited for the task of flow feature extraction. 
        However, adapting these algorithms to physical flows presents some uniques challenges. 
        
        The first challenge is in representing the flow in a manner conductive to the application of these algorithms. 
        An image is represented with three invariant (i.e. not dependent on choice of coordinates) RGB color values at each pixel. 
        Similarly, we represent the flow by a suitable list of invariant physical properties at each of a number of discrete points. 
        The second challenge is that while algorithms developed to use images require inputs with rectangular domains, the physical domain of fluid flows is generally not rectangular, containing boundaries of arbitrary shape as well as holes in the domain were solid objects exist. 
        To circumvent this challenge, in this initial study flows were restricted to flows with singly-connected domains, and were mapped to rectangular domains. 
        The last challenge is that while objects in images can be identified using rectangular bounding boxes of moderate aspect ratios that mostly contain only the object in question, not all flow features can be identified in this manner. 
        To address this we distinguish between what we call \emph{discrete features} such as shed vortices which can be encompassed by a bounding box and \emph{continuous features} such as boundary layers which cannot. 
        Continuous feature problems have no analog to image recognition and the approach taken was to provide a boolean field of points that belong to that feature, rather than a rectangular bounding box. 

        One final note on nomenclature: this paper combines concepts from different fields and there is a terminology conflict that must be addressed. 
        In image recognition the goal is to detect objects within the image, and this is done by first identifying smaller scale patterns (e.g. horizontal lines, or sharp gradients) referred to as \emph{features}. 
        The conflict arises in that in fluid flows the `objects' to be detected are also referred to as \emph{features}. 
        We will avoid ambiguity by referring to these as \emph{local-features} and \emph{flow features} respectively. 
        The term \emph{feature} is also commonly used to describe the point-based properties of the flow, but these will be referred to as input parameters.

\section{Methodology}
    \label{S:methodology}
    The methodology, which we we will refer to as Fluid R-CNN, is adapted from the R-CNN method~\cite{girshick_rich_2014,ren_faster_2015} to work for fluid flow feature identification. 
    The Fluid R-CNN can be divided into four steps. 
    First, the flow is represented as a three-dimensional rectangular grid, with each point described by a list of invariant physical flow properties, analogous to the three RGB values used to describe pixels in images. 
    Next, in the region proposal step, subsets of the flow are selected for classification. 
    In the classification step, the proposed regions are evaluated using a trained CNN and classified as being either an instance of the flow feature of interest or as background. 
    Finally, in the selection step, the information from all the region evaluations is used to select and report the number and location of flow features. 

    \subsection{Input Array}
        The starting point of the method are fluid flow fields such as those obtained from either mean or time-resolved simulations or experimental data. 
        In particular for the case studies presented here they are mean fields from Reynold-averaged Navier-Stokes (RANS) simulations. 
        Following the procedures in Wang et al.~\cite{wang_physics-informed_2017} and Ling et al.~\cite{ling_evaluation_2015}, ten input parameters are used to characterize the flow at each cell. 
        These parameters are non-dimensional, Galilean invariant, and based only on the flow properties at that point. 
        The ten input parameters are summarized in \ref{app:A} and include turbulence intensity, non-orthogonality of velocity and its gradient, and streamline curvature.

        In these test cases we restrict our inputs to flows with a simply-connected domain and uniform mesh.
        The uniform mesh can be directly represented as a rectangular three-dimensional array, with each element corresponding to a mesh cell and having associated $(x,y,z)$ coordinates, cell volume, and 10-dimensional input.
        This array representation of the flow fields is the input to the Fluid R-CNN method, and is referred to as the \emph{input array}.

    \subsection{Region Proposal}
        While many region proposal algorithms are being developed \cite{girshick_rich_2014} that can efficiently produce proposals for objects in thousands of classes, a brute search approach was used.
        The brute search consists of searching the input array with a number of windows of different shapes, using specified stride length.
        All the resulting regions are considered possible flow features to be classified by the CNN. 
        Brute search does not add significant computational cost in the test cases presented since only a single flow feature is being searched for (binary classification). 
        The advantage of more sophisticated region proposal algorithms is more evident when a large number of classes are considered.  

\begin{figure}[htb]
        \centering
        \subfloat[Rectangular subregion of the input array.]{
            \includegraphics[width=0.45\textwidth,valign=c]{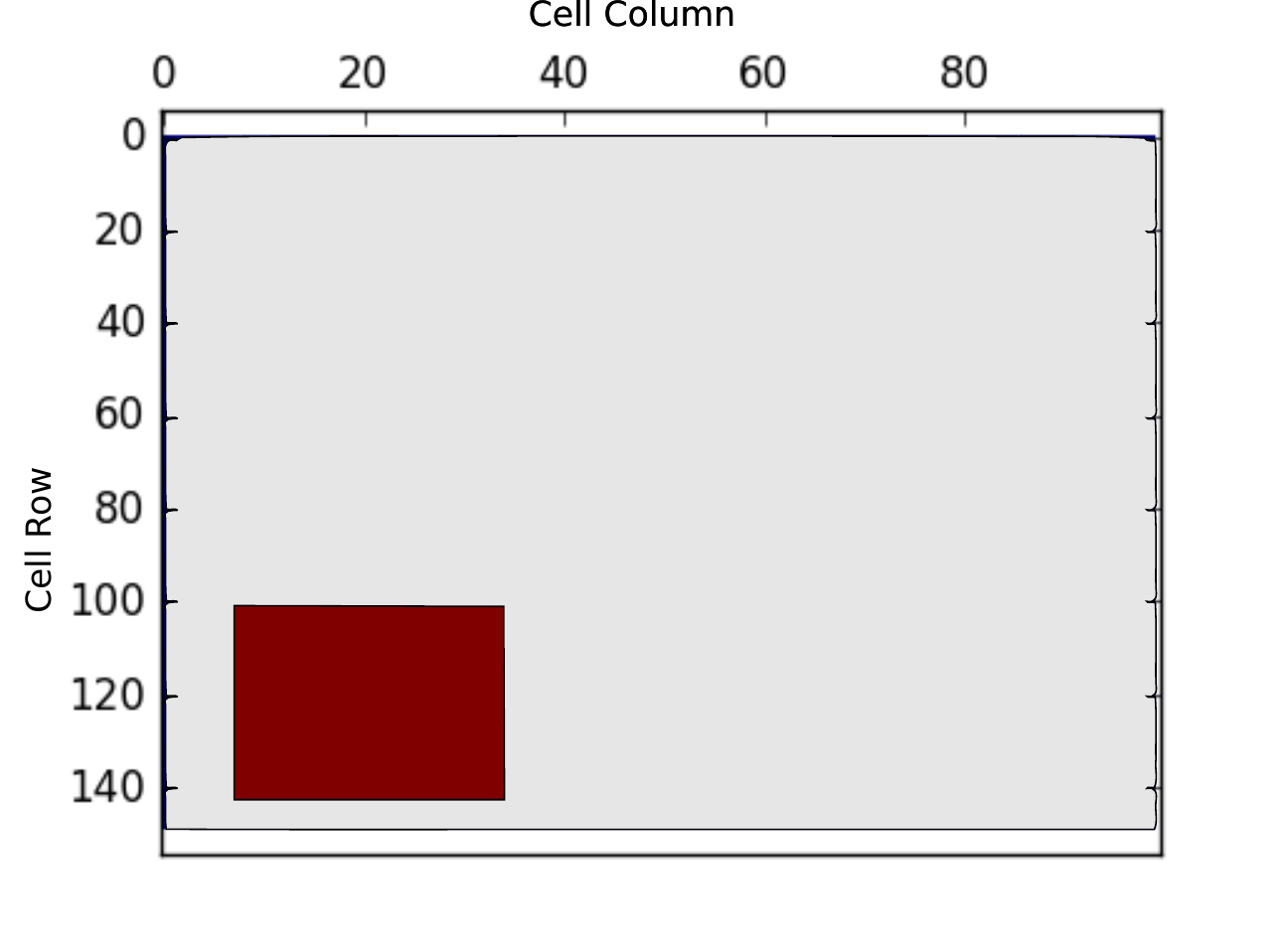}
            \label{fig:dmap_m}
        } \quad
        \subfloat[Same region in the physical domain.]{
            \includegraphics[width=0.45\textwidth,valign=c]{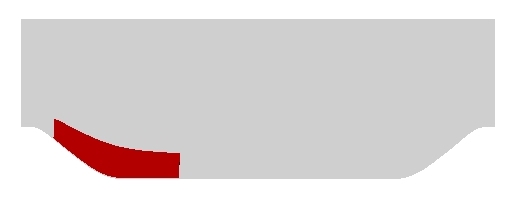}
            \label{fig:dmap_p}
            \vphantom{\includegraphics[width=0.45\textwidth,valign=c]{ph_recirc_region_map.pdf}}
        }
        \caption{Example of a rectangular sub-region of the input array (a), and its corresponding region in the physical domain (b). This example corresponds to the human-labeled recirculation region for the 2D periodic hills flow.}
        \label{fig:dmap}
        \end{figure}

        A consequence of scanning a rectangular representation of the flow is that since elements of the input array correspond to cells with different volumes, a scanning window of fixed dimensions (e.g. $n \times m$ cells) will result in regions of vastly different physical shapes and dimensions. 
        Figure~\ref{fig:dmap} shows a rectangular subregion of the input array, and its corresponding region in the physical domain. 
        A second consideration is that the CNN only accepts rectangular inputs of a fixed size, while the region proposal selects regions of different sizes and aspect ratios. 
        This second problem is shared with object detection in images, where region proposals might have different size and aspect ratios than the CNN input, and is solved by warping the image and using a CNN trained with warped images. 

\begin{figure}[htb]
          \centering
          \subfloat[Region as CNN input of shape $24\times24$.]{
            \includegraphics[width=0.4\textwidth,valign=c]{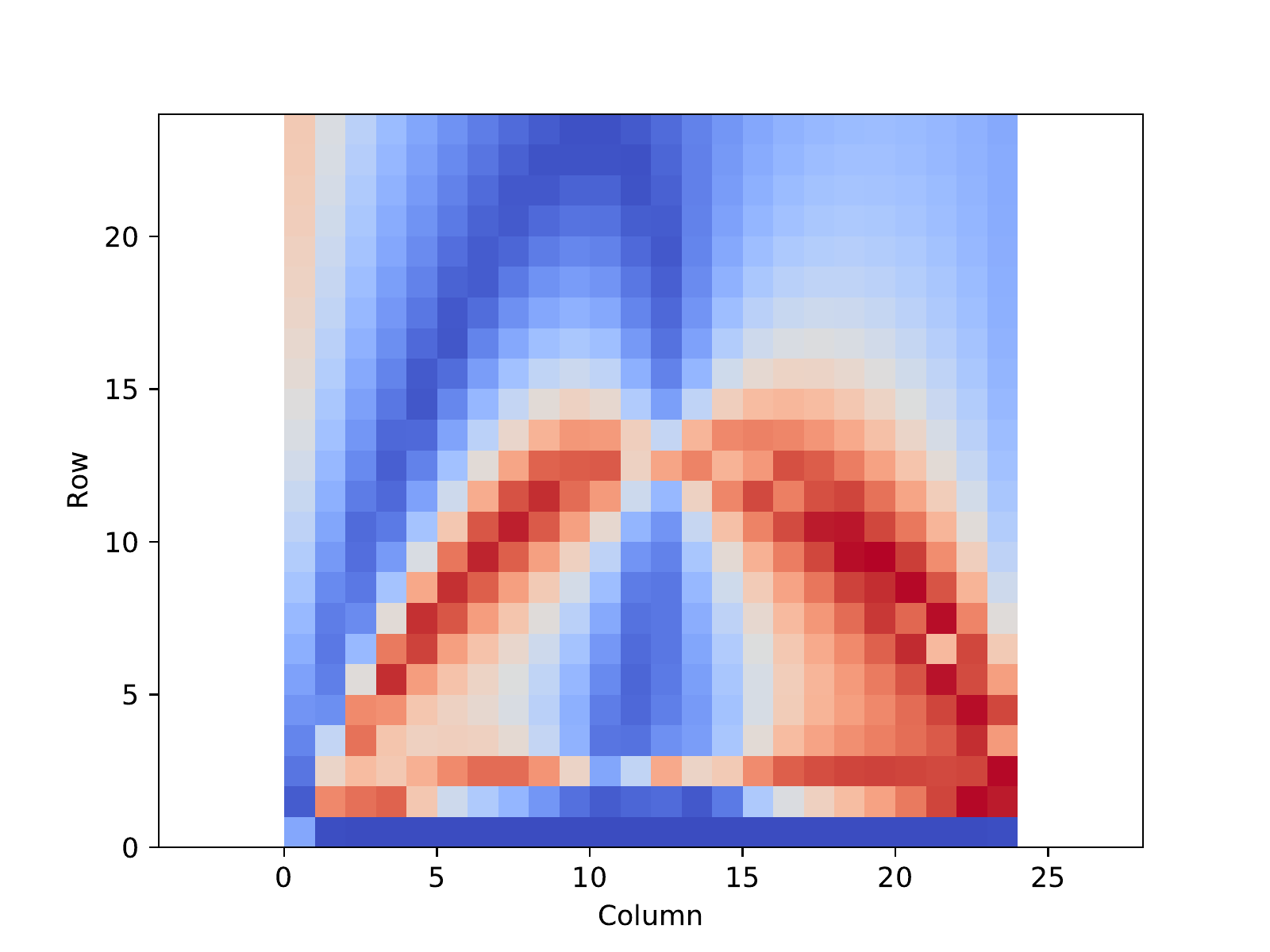}
            \label{fig:warp_m}
            } \quad
          \subfloat[Region in the physical domain.]{
            \includegraphics[width=0.5\textwidth,valign=c]{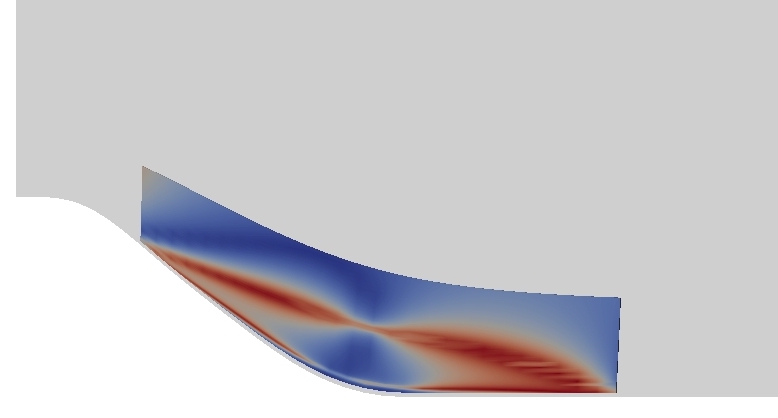}
            \vphantom{\includegraphics[width=0.4\textwidth,valign=c]{UGradMis_map.pdf}}
            \label{fig:warp_p}
            }
          \caption{Mapping of a proposed region in the physical domain (b) to CNN input (a). The example region corresponds to the region shown in Figure~\ref{fig:dmap}, and the input shown is the non-orthogonality of velocity and its gradient.}
          \label{fig:warp}
        \end{figure}

        The goal is then to map each proposed region into a rectangular region of the specified size, with as little deformation as possible. 
        This was done by first normalizing each element in $[:,j,k]$ rows based on the cells' $x$-coordinates, each element in  $[i,:,k]$ rows based on the cells' $y$-coordinates, and each element in $[i,j,:]$ rows based on the cells' $z$-coordinates. 
        This, now rectangular, region is interpolated to a uniform grid with the required CNN input dimensions. 
        The process is illustrated in Figure~\ref{fig:warp}. 

    \subsection{Classification Using CNN}
        Following the region proposal step, each proposed region must be classified.
        The classification step is done using a convolutional neural network, trained to differentiate between background regions and regions containing the feature of interest.
        The first step in creating a CNN is determining its architecture.
        The architecture must be flexible enough to make the classification task possible, while being simple enough to keep the number of training parameters and computational cost low.
        The second step is training the CNN, in which the optimal values for the trainable parameters are learned based on the training examples provided.

        \subsubsection{CNN Architecture}
            \label{ss:CNN}
\begin{figure}[htb]
              \centering
              \includegraphics[width=0.8\textwidth]{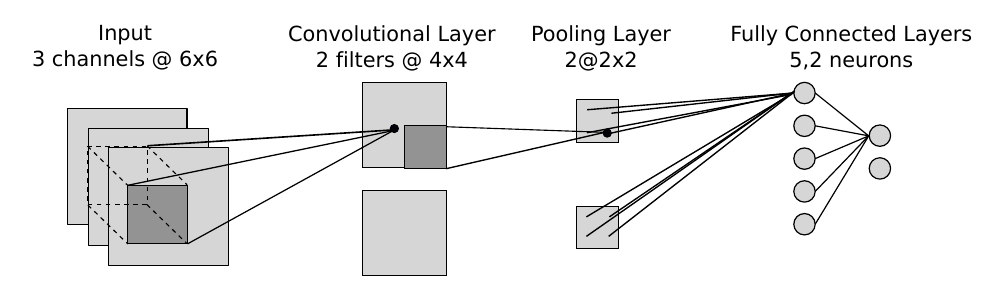}
              \caption{Example schematic of the CNN architecture with one convolutional layer with 2 feature maps and a pooling layer, followed by two dense layers, the first with 5 linear rectified neurons, and the output layer with 2 softmax neurons. For clarity of illustration, the input layer is shown as having two physical dimensions rather than all three. For each layer following the input layer, the schematic shows the input for a single value in that layer.}
              \label{fig:cnn_arch}
            \end{figure}

            The CNN architecture was designed with the goal of obtaining as simple a network as possible, while still being able to correctly identify the features of interest. 
            All cases in this study use a similar CNN, with a single convolutional layer, a pooling layer, and an output layer consisting of 2 softmax neurons. 
            A schematic of the CNN architecture used in this study is shown in Figure~\ref{fig:cnn_arch}. 
            For clarity of illustration, Figure~\ref{fig:cnn_arch} shows an input with only two physical dimensions rather than the full 3 dimensions actually used.

            The input layer corresponds to a region in the flow domain, with each input point described by a $10$-dimensional input vector ($10$ channels). 
            This input layer is a tensor of size $N_c \times N_x \times N_y \times N_z$, where $N_c$ is the number of channels and $N_x$, $N_y$, $N_z$ are the number of points in each spatial dimension. 
            The input layer is followed by a convolutional layer, which consists of $N_f$ local-feature maps. 
            Each map contains information of the presence of a specific local-feature (e.g. a sharp gradient, or a horizontal line) at different locations in the input. 
            The map is created by sliding  a window through the spatial dimensions of the input layer, outputting a single neuron activation value at each location considered. 
            The sliding window over the inputs is called the \emph{local receptive field} and has dimensions $N_c \times N_{fx} \times N_{fy} \times N_{fz}$. 
            The neuron activation chosen is the rectified linear unit (ReLU), with activation given by 

            \begin{equation}
            \label{eq:ReLU}
            y = \max(0,\sum_{i=1}^{N_{rf}} ( w_i x_i) + b,
            \end{equation}
            where $N_{rf}$ and $x_i$ are the number of inputs and the individual elements in the local receptive field respectively, and $w_i$ and $b$ are the \emph{weights} and \emph{bias} respectively. 
            The weights and the bias are parameters that need to be determined through training. 
            These weights and bias are the same for all spatial locations, i.e. when the local receptive field is moved to a different location in the input the weights and bias are unchanged. 
            This enforces translational invariance, greatly reducing the number of trainable parameters. 
            The example in Figure~\ref{fig:cnn_arch} has a 2-dimensional input of size $6\times6$ with 3 channels ($N_c=3$, $N_x=6$, $N_y=6$). 
            The input layer is followed by a convolutional layer with two filters, each with a local receptive field of shape $3\times3\times3$ ($N_c=3$, $N_{fx}=3$, $N_{fy}=3$) and stride length of one in both directions, resulting in the local feature maps of size $4\times4$. 

            Immediately following the convolutional layer is a pooling layer, which has the role of condensing the information and reducing the size of the input.  
            It does this by dividing the feature maps into windows of specified size (stride length is window dimensions) and summarizing each region with a single value. 
            Max pooling was used for this purpose, where the output is the maximum value in the window. 
            This results on information on whether the feature is found \emph{anywhere} in that window. 
            These could potentially be followed by more convolution-pooling layer pairs with the goal of identifying higher scale features from smaller scale features, but for simplicity only one convolutional layer was used. 
            The example in Figure~\ref{fig:cnn_arch} has a max pooling layer with input of size ($2\times2$), which condenses each feature map (size $4\times4$) to size $2\times2$. 

            Finally, the CNN has one or more layers of fully connected neurons, with the output layer being a \emph{softmax} layer of size equal to the number of categories being considered.  
            Each neuron in a softmax layer outputs a probability of the input belonging to that category, with the sum of all the outputs equal to one. 
            The output of the softmax neuron is given by  
            
            \begin{equation}
            y_i = \frac{e^{z_i}}{\sum_{j=1}^{N_{curr}}e^{z_j}},
            \end{equation}
            
            \begin{equation}
            z_j = \sum_{k=1}^{N_{prev}} ( w_{k,j} x_{k}) + b,
            \end{equation}
            where $N_{curr}$ and $N_{prev}$ are the number of neurons in the current (softmax) and previous layers respectively, and $x_k$ and $w_{k,j}$ are the output and weight from the $k$\textsuperscript{th} neuron of the previous layer to the $j$\textsuperscript{th} neuron of the current layer. 
            In this study ReLU neurons are used for all intermediate layers and two softmax neurons are used as the final layer, classifying the input as belonging to the feature of interest or to the background. 
            For all cases the weights were initialized by sampling a normal distribution with mean zero and standard deviation $0.1$, and all biases were initialized to zero. 

        \subsubsection{CNN Training}
            \label{ss:train}
            The first step in training the CNN is creating examples of regions that are instances of the flow feature of interest and examples of regions that belong to the background. 
            For discrete features the training inputs are the entirety of the flow field with the instances of the flow features of interest labeled. 
            Labeling is done by identifying a bounding rectangle that completely encloses the flow feature of interest.
            Each labeled region can provide many training examples by shifting the bounding box and considering partial overlap.
            Using different window sizes, the training case is scanned to create potential training examples. 
            For each potential example, the maximum intersection over union (IoU) with a labeled region is calculated based on cell volume. 
            The IoU for two regions $A$ and $B$ is given by 
            
            \begin{equation}
            IoU(A,B) = \frac{A\cap B}{A\cup B},
            \end{equation}
            where $A\cap B$ and $A\cup B$ denote the volume of their intersection and union respectively. 
            Regions with IoU above a certain threshold (e.g. 50\%) are considered examples of the flow feature. 
            Those below the threshold are considered background. 
            Furthermore those in the background that have an IoU larger than some second threshold (e.g. 25\%) are considered difficult examples \cite{girshick_rich_2014} of the background. 
            These regions are then down-sampled to the desired number of training cases, ensuring sufficient examples of each category are retained. 

            The CNN is then trained, with the data prepared as above. 
            Training consists of optimizing the CNN parameters, i.e. the biases associated with each neuron and the weights associated with each connection. 
            The training is done using stochastic gradient descent and back-propagation, with a categorical cross-entropy cost function. 
            No dropout or regularization was used. 
            Using gradient descent, each parameter $a_i$ (weights and biases) is updated as 
            
            \begin{equation}
            a_i^{(n+1)} = a_i^{(n)} - \gamma \nabla C(a_i),
            \end{equation}
            where the superscript indicates the update step, the learning rate $\gamma$ is a hyper-parameter, and $C$ is the cost function which depends on all the parameters $a_i$ and all the training examples. 
            The categorical cross-entropy cost function is given by  
            
            \begin{equation}
            C = -\frac{1}{N_t}\sum_{j=1}^{N_t}\sum_{i=1}^{N_{cat}}t_{i,j}\ln{(p_{i,j})},
            \end{equation}
            where $t_{i,j}$ and $p_{i,j}$ are the true and predicted values of the $i$\textsuperscript{th} output neuron for the $j$\textsuperscript{th} training example, $N_{cat}$ is the number of categories (i.e. number of neurons in last layer), and $N_t$ is the number of training examples. 
            For a given training case ($j$) one $t_{i,j}$ has a value of one and all others have a value of zero. 
            Stochastic gradient descent refers to approximating the gradient of $C$ using only a random subset (mini-batch) of the training data at each update step. 
            Back-propagation is a method to calculate the gradients of $C$ with respect to each weight and bias by applying the chain rule to iteratively obtain the gradients at each layer, starting from the output layer. 

            At each training iteration a mini-batch is created with a random selection of a specified number of training examples. 
            Each mini-batch is forced to contain a certain proportion of each type of examples (e.g. $50\%$ easy background examples, $25\%$ difficult background examples, and $25\%$ flow feature examples). 
            This is to ensure the training sees enough examples of the flow feature and difficult background examples.  
            The training goes on until an acceptable convergence is seen in the cost function. 

            The training for continuous features is similar except that section cuts of the original flow are used as the training cases, rather than the entire flow domain. 
            The flow feature of interest is only identified and labeled in these section cuts. 
            This procedure was chosen because for continuous features a single bounding box mostly containing the flow feature is not possible in the whole domain, but is possible in section cuts of the domain. 

    \subsection{Selection}
\begin{figure}[htb]
         \centering
         \includegraphics[width=0.6\textwidth]{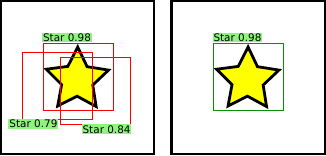}
         \caption{Example of how non-maximum suppression is used to select a single region for a given flow feature. The region proposal and CNN evaluation might result in a number of regions above the classification threshold (left). By eliminating any region with intersection with another region with higher probability, a single region is selected (right) to represent that particular flow feature.}
         \label{fig:select}
        \end{figure}

        The goal of the selection step is report one single region per flow feature based on the results from the classification step. 
        Each region from the region proposal step is classified by the CNN as either background or flow feature.  
        For discrete feature problems, many such regions describe the same feature. 
        This is specially the case since the CNN is trained to positively classify partial features (typically as low as 50\% overlap), and some of the cell volumes are very small. 
        Because of this, two consecutive regions might be extremely similar if the stride length is small. 
        The goal is to report one single region per flow feature. 
        This is achieved using non-maximum suppression, rejecting any region that overlaps another region with higher probability of being a feature, as illustrated in Figure~\ref{fig:select}. 
        For \emph{testing} cases, a successful identification is considered to be a selected region with with IoU with a human labeled region above a certain threshold (e.g. typically $IoU \geq 0.5$ for object detection in images). 
        The accuracy of the method can then be quantified by the percentage of regions that were correctly identified, as well as the percentage of false positives. 

        For continuous feature problems there are no discrete regions to be identified, but rather a continuum. 
        For these cases selection was done by assigning a boolean value to each cell, with a \emph{True} value for any cell that is within at least one of the regions classified by the CNN as being a feature. 
        The selected region is then the union of all regions classified positively by the CNN. 
        Since non-maximum suppression is not used for continuous feature problems, more aggressive choice of thresholds is needed to avoid selected regions much larger than the flow feature. 
        In the context of the \emph{feature extraction pipeline} presented in Post et al.~\cite{post_state_2003} the presented method performs the selection step (i.e. identifying all points that are part of the feature of interest) for both types of problems. 
        It also performs the clustering step (i.e. dividing these points into discrete instances of the feature of interest) for the discrete feature problems but not for the continuous feature problems. 

    \subsection{Implementation}
        The method was implemented in Python and runs in CPUs. 
        We used the Lasagne~\cite{lasagne_short} and Theano~\cite{theano_short} libraries to create the CNN. 
        The data creation task was parallelized with each training case evaluated independently. 
        The CNN classification step was also parallelized with different batches of proposed regions sent to different CPUs. 
        The training step was done in serial. 
        As an example, Table~\ref{tb:compcost_c1} shows the computational cost and parallelization for the first case study.
        The largest costs are associated with the creation of training data (including the warping and interpolation) and the evaluation of the CNN. 
        While some care was used to improve the cost, this was not the goal of this study and it can still be improved significantly. 
        In image recognition there has been a push towards real-time identification, which is being achieved through algorithmic choices and GPU implementation. 
        Some of the same techniques can be leveraged to speed-up our implementation.

        \iftoggle{extra_spacing}
        {
            \renewcommand{\baselinestretch}{1.0}
        }{}
        
        \begin{table}[!ht]
        \centering
        \caption{Summary of computational cost for Case Study I.}
        \label{tb:compcost_c1}
        \begin{tabular}{ c|c|c } 
        Task & No. CPUs & time [$s$] \\ 
        \hline
        Create training data (3 train cases) & $3$  & $1,069$ \\
        Train (500 steps)                    & $1$  & $169$ \\
        Evaluate regions with CNN            & $72$ & $1,982$ \\
        \hline
        \end{tabular}
        \end{table}

        \iftoggle{extra_spacing}
        {
            \renewcommand{\baselinestretch}{\stretchval}
        }{}

\section{Results}
    \label{S:results}
    Three case studies were investigated, using data from Reynolds-averaged Navier-Stokes (RANS) simulations. 
    The first is a discrete feature problem: identifying the recirculation region in a 2D flow (Figure~\ref{fig:c1_res}). 
    The other two are continuous feature problems: identifying the boundary layer in 2D flows (Figure~\ref{fig:c2_res}), and a horseshoe vortex in a 3D flow (Figure~\ref{fig:c3_io}). 

    \subsection{Case Study I - Identifying 2D Recirculation Region}
\begin{figure}[htb]
          \centering
          \subfloat[Curved backwards-facing step.]{
            \includegraphics[width=0.30\textwidth,valign=c]{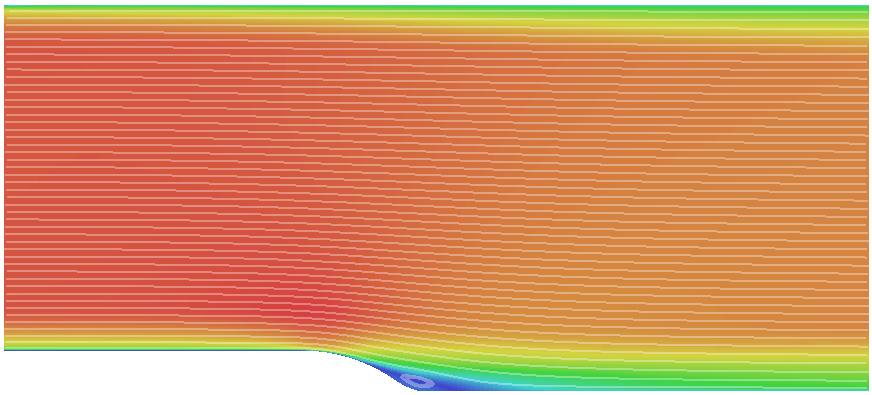}
            \label{fig:rans_cbs}
            } \quad 
          \subfloat[Convergent-divergent channel.]{
            \includegraphics[width=0.45\textwidth,valign=c]{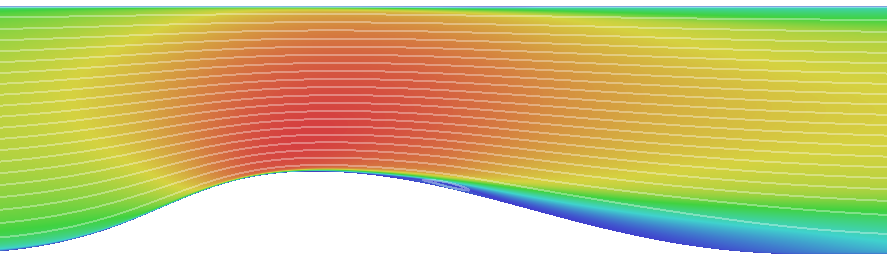}
            \vphantom{\includegraphics[width=0.30\textwidth,valign=c]{Ucontour_CBFS.png}}
            \label{fig:rans_cdc}
            } \\
          \subfloat[Periodic hills.]{
            \includegraphics[width=0.45\textwidth,valign=c]{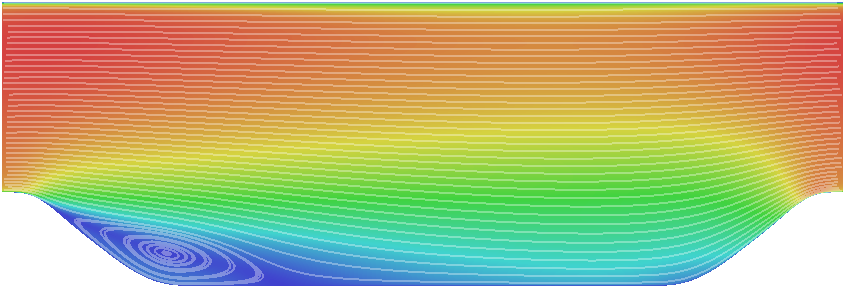}
            \vphantom{\includegraphics[width=0.15\textwidth,valign=c]{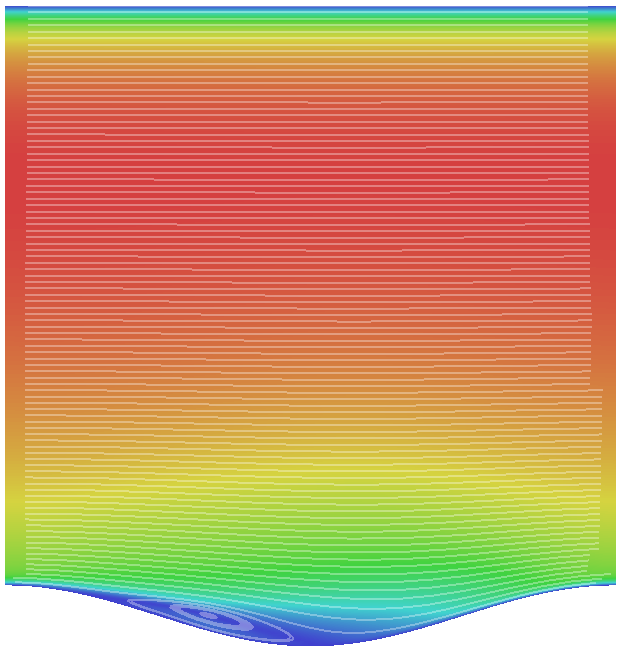}}
            \label{fig:rans_ph}
            } \quad 
          \subfloat[Wavy channel.]{
            \makebox[0.2\textwidth][c]{
            \includegraphics[width=0.15\textwidth,valign=c]{Ucontour_WC.png} }
            \label{fig:rans_w}
            }
          \caption{Simulation cases used in Case Study I and II. The images show velocity magnitude (color map with blue and red denoting the smallest and largest magnitudes) as well as velocity streamlines.}
          \label{fig:rans}
        \end{figure}

        The first case study involved identifying regions of recirculation in 2D flows. 
        This is a discrete feature problem, in which the CNN is trained using flow cases with the whole recirculation region labeled. 
        The CNN was trained using three flows: periodic hills, wavy channel, and curved backwards-facing step. 
        Each of these cases contained a single recirculation region, identified visually based on looping streamlines. 
        The human-labeled recirculation region is shown for one of the training cases in Figure~\ref{fig:dmap}.
        The methodology was then tested on a new flow: a convergent-divergent channel. 
        These cases are summarized in Table~\ref{tb:rans} and shown in Figure~\ref{fig:rans}. 
        The Reynolds number is based on hill/step height and bulk mean velocity at the hill crest. 
        For all cases the RANS turbulence model used was the Launder-Sharma $k-\epsilon$.
        The convergent-divergent channel was chosen as the test case since it is thought to contain the most different recirculation region, making it the most difficult one to predict. 
        The most noticeable difference is the long aspect ratio and short height of the recirculation region. 

        \iftoggle{extra_spacing}
        {
            \renewcommand{\baselinestretch}{1.0}
        }{}

        \begin{table}[!ht]
        \centering
        \caption{Summary of the four RANS cases used in Case Study I and II.}
        \label{tb:rans}
        \begin{tabular}{ c|c|c } 
        Case & Reynolds Number & Mesh Size  \\
         & & ($rows \times columns$) \\
        \hline
        Curved backwards-facing step & 13700 & $199\times199$  \\ 
        Convergent-divergent channel & 12600 & $499\times192$ \\ 
        Periodic hills & 10595 & $99\times149$ \\ 
        Wavy channel & 6760 & $50\times160$ \\ 
        \hline
        \end{tabular}
        \end{table}

        \iftoggle{extra_spacing}
        {
            \renewcommand{\baselinestretch}{\stretchval}
        }{}

        The CNN architecture consists of inputs size $24\times24$, ten feature maps in the convolutional layer with window size $5\times5$ and max pooling with $2\times2$ windows, and two dense layers of $10$ ReLU and $2$ softmax neurons consecutively.   
        The CNN was trained for $500$ training steps with a learning rate of $0.01$ using mini-batches containing $5$ recirculation cases, $5$ difficult background cases, and $10$ background cases.  
        The region proposal was done with $15$ window sizes corresponding to aspect ratios of $1:1,\ 2:1,\ 1:2$ and scale factor of $10,\ 20,\ 30,\ 40,\ 50$ cells.
        A stride of $2$ cells was used in each direction, resulting in $266,525$ proposed regions. 

\begin{figure}[htb]
          \centering
          \subfloat[Input array.]{
            \includegraphics[width=0.5\textwidth,valign=c]{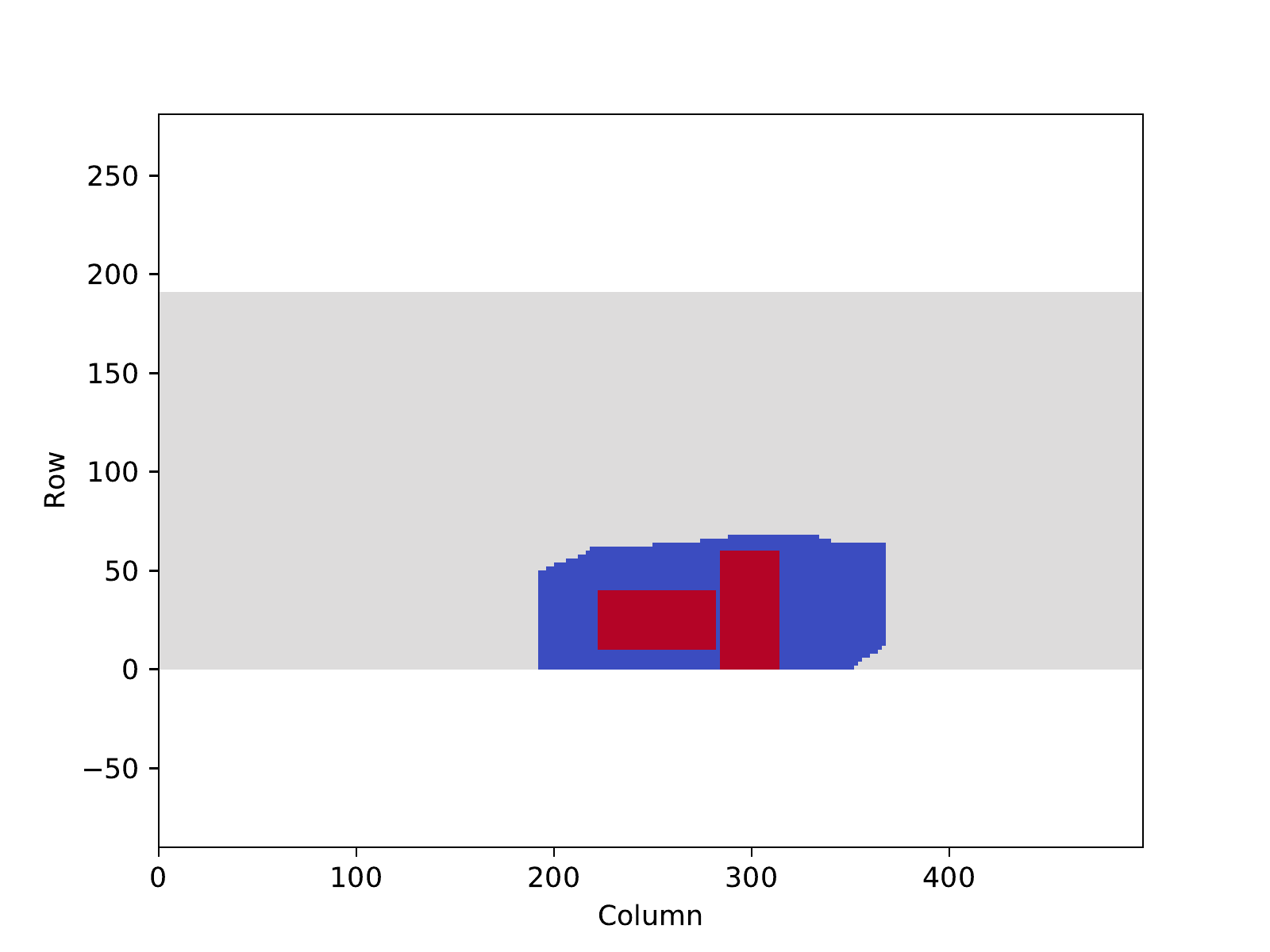}
            \label{fig:c1_res_m}
            } \\
          \subfloat[Physical domain.]{
            \includegraphics[width=0.75\textwidth,valign=c]{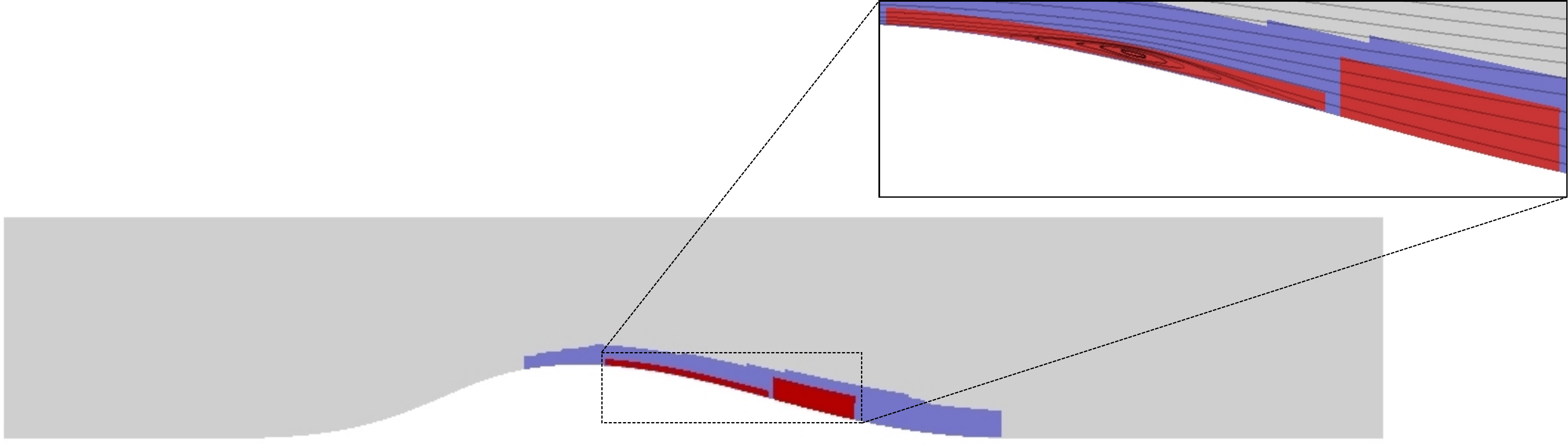}
            \label{fig:c1_res_p}
            }
          \caption{Case Study I results: recirculation regions in the convergent-divergent channel flow. Blue denotes regions that the CNN classified as recirculation, and red denotes those regions selected as instances of recirculation.}
          \label{fig:c1_res}
        \end{figure}

        The Fluid R-CNN resulted in two regions selected as instances of recirculation, out of $2,242$ evaluated positively by the CNN. 
        One region was correctly selected, with a $98\%$ probability of being recirculation, and an intersection over union of $68\%$ with the human labeled region. 
        The other region, a false positive, was assigned a $77\%$ probability of being recirculation and has no intersection with any human labeled region. 
        This false positive is likely due to the limited amount of training cases, which cannot provide enough training examples of all possible background regions. 
        In the absence of large quantities of training data, a probability threshold can be manually tuned to a value higher than $50\%$ to reduce these false positives.  
        The results are summarized in Table~\ref{tb:c1-results} and shown in Figure~\ref{fig:c1_res}. 
        Figure~\ref{fig:c1_res_p} also shows a close-up view of the streamlines in the two selected regions.
        
        \iftoggle{extra_spacing}
        {
            \renewcommand{\baselinestretch}{1.0}
        }{}

        \begin{table}[!ht]
        \centering
        \caption{Case Study I results: list of selected regions with assigned probability of being a recirculation region and IoU with a correct, human-labeled, region.}
        \label{tb:c1-results}
        \begin{tabular}{ c|c|c } 
        Region & Probability & IoU with correct region\\
        \hline
        1 & $0.98$ & $0.68$\\
        2 & $0.77$ & $0.00$\\
        \hline
        \end{tabular}
        \end{table}

        \iftoggle{extra_spacing}
        {
            \renewcommand{\baselinestretch}{\stretchval}
        }{}

    \subsection{Case Study II - Identifying 2D Boundary layer}
\begin{figure}[htb]
          \centering
          \includegraphics[width=0.5\textwidth]{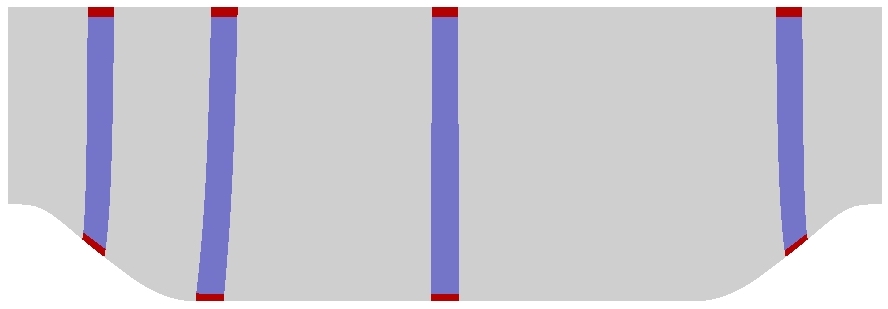}
          \caption{Section cuts (blue) and human-labeled boundary layer regions (red) used to train the boundary layer CNN for Case Study II.}
          \label{fig:c2_cases}
        \end{figure}

        For the second case study the boundary layer, a continuous feature, is identified in two different flows. 
        The flows used are the periodic hills, and the convergent-divergent channel flows shown in Figure~\ref{fig:rans}. 
        A CNN was trained using four section cuts from the periodic hills case, shown in Figure~\ref{fig:c2_cases}. 
        The boundary layer was identified visually based on wall-distance Reynolds number. 

\begin{figure}[htb]
          \centering
          \subfloat[Periodic hills.]{
            \includegraphics[width=0.35\textwidth,valign=c]{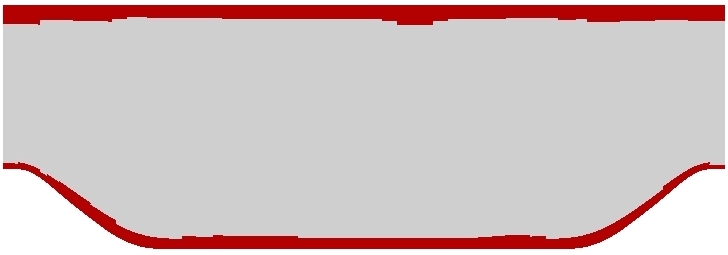}
            \label{fig:c2_res_ph}
            } \quad
          \subfloat[Convergent-divergent channel.]{
            \includegraphics[width=0.55\textwidth,valign=c]{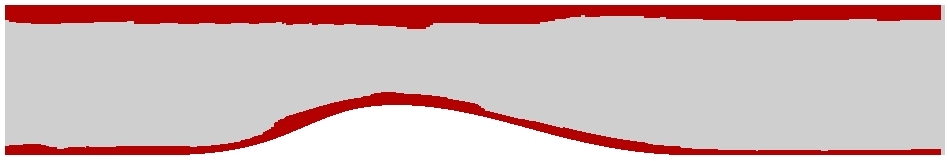}
            \vphantom{\includegraphics[width=0.35\textwidth,valign=c]{ph_bl_flow.jpg}}
            \label{fig:c2_res_cdc}
          }
          \caption{Case Study II results: boundary layers (red/dark gray) in the periodic hills and convergent-divergent channel flows.}
          \label{fig:c2_res}
        \end{figure}

        The CNN architecture was identical to that for Case Study I. 
        The CNN was trained for $500$ training steps with a learning rate of $0.01$ using mini-batches containing $5$ boundary layer cases, no difficult background cases, and $15$ background cases.  
        All training windows were restricted to be the same width as the section cuts, namely $3$ cells. 
        The Fluid R-CNN was first tested on the entire periodic hill case, to identify the entire boundary layer after training with sectional cuts from the same case. 
        The region proposal for the periodic hills case was done with 5 windows of width $3$ and heights ${10,15,20,25,30}$ cells with stride of $2$ in both directions, resulting in $15,974$ total regions.
        The results are shown in Figure~\ref{fig:c2_res_ph}.  
        
        The Fluid R-CNN was then used on a new flow, the convergent-divergent channel, for which the CNN training saw no data from.
        The region proposal for the convergent-divergent channel case was done with 6 windows of width $3$ and heights ${10,20,30,40,50,60}$ cells with stride of $2$ in both directions, resulting in $15,974$ total regions.
        The results are shown in Figure~\ref{fig:c2_res_cdc}.  

        For both cases the algorithm was able to identify the entirety of the boundary layer.  
        This was verified visually rather than quantitatively. 
        One noticeable quirk of the results is that the boundary layer is significantly thicker than the human labeled examples used for training. 
        This is due to the fact that the creation of training cases considers partial overlap regions, up to some IoU threshold, as examples of boundary layer. 
        For this reason the IoU threshold was increased to $0.75$ (from the typical $0.5$). 
        For this same reason and because the boundary layer is a gradient with no clear edge, no difficult examples were used for background and its threshold was lowered to $0.1$ from the typical $0.25$. 
        An alternative would be to use training examples with intermediate values of the probabilities for the different categories, rather than with examples labeled as $100\%$ probability of belonging to one or the other. 

    \subsection{Case Study III - Identifying 3D Horseshoe Vortex}
\begin{figure}[htbp]
          \centering
          \subfloat[Schematic.]{
            \includegraphics[width=0.45\textwidth,valign=c]{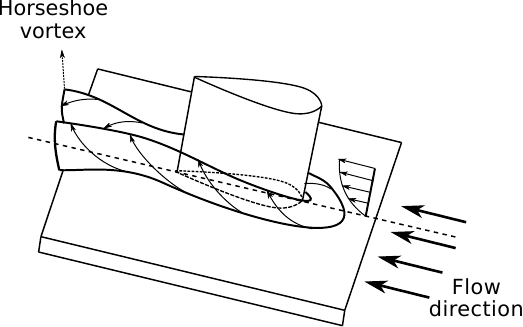}
            \label{fig:c3_schem}
            } \quad
          \subfloat[Results.]{
            \includegraphics[width=0.45\textwidth,valign=c]{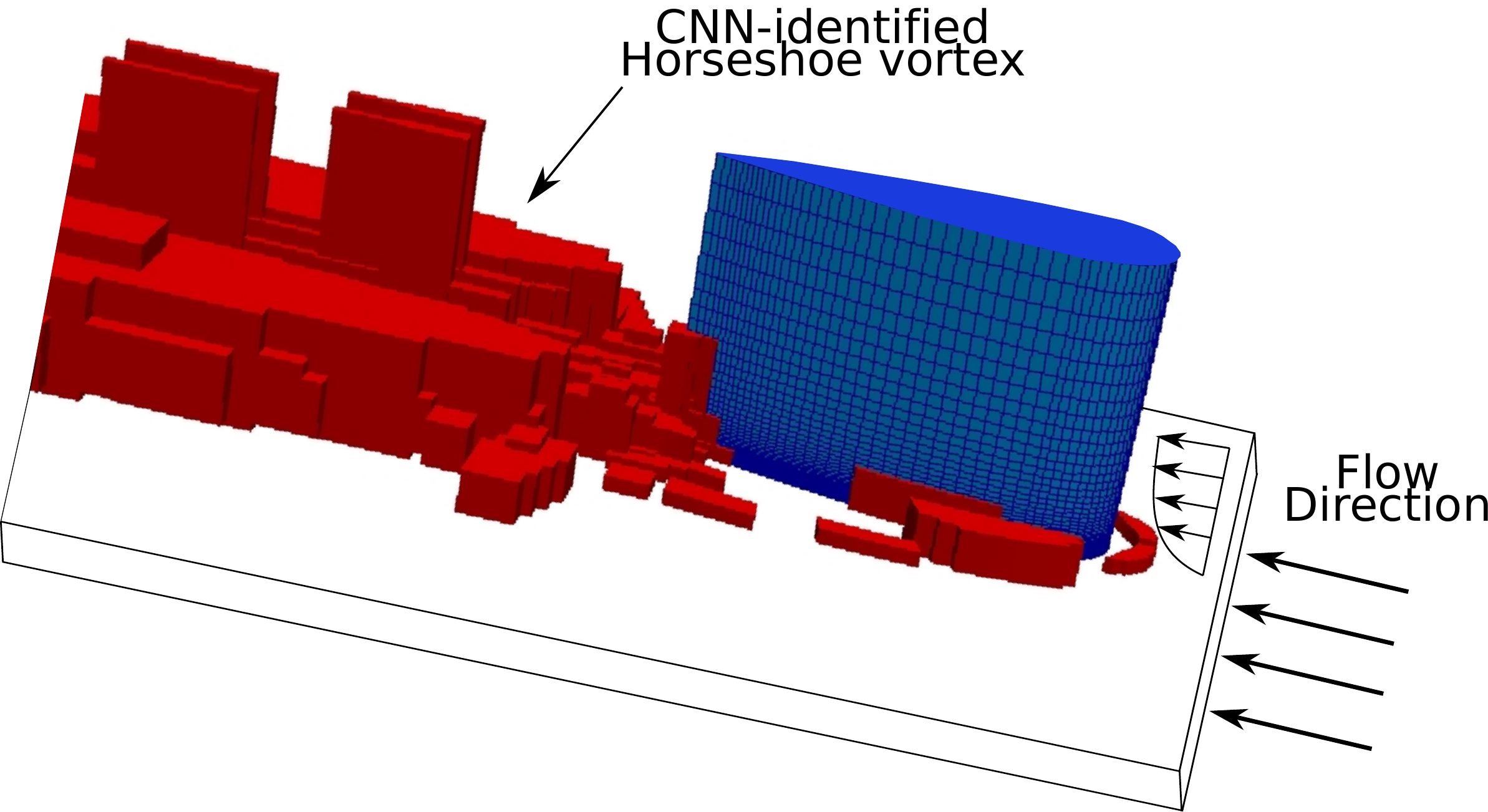}
            \label{fig:c3_diag}
            }
          \caption{Schematic of a horseshoe vortex around the wing-body junction geometry used in Case Study III (a), and results from the Fluid R-CNN method (b).}
          \label{fig:c3_rans}
        \end{figure}

        The third case study consists of identifying the horseshoe vortex in a 3D wing-body junction flow based on the experimental setup of Devenport et al.~\cite{devenport_simpson_1990}.
        The problem is illustrated in Figure~\ref{fig:c3_rans}. 
        This is a continuous feature problem and the CNN was trained with section cuts of the domain. 
        The Fluid R-CNN was then tested in the same flow in order to identify the entirety of the horseshoe vortex. 
        The horseshoe vortex identification was done in only half the domain, but the results are mirrored for visualization. 
        The RANS simulation for this case was done for half of the domain, using a symmetry plane, and a mesh size of $149\times98\times49$ rows, columns, and layers respectively.
        The Reynolds number is about $10^5$ based on airfoil thickness, and the turbulence model used was the $k-\omega\ SST$ model.
        
\begin{figure}[htbp]
          \centering
          \subfloat[Training cases.]{
            \includegraphics[width=0.45\textwidth,valign=c]{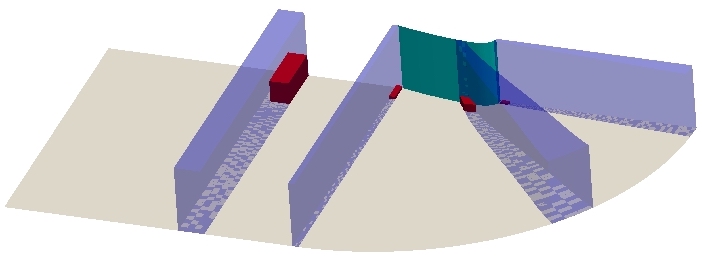}
            \label{fig:c3_train}
            } \quad
          \subfloat[Identified horseshoe vortex.]{
            \includegraphics[width=0.45\textwidth,valign=c]{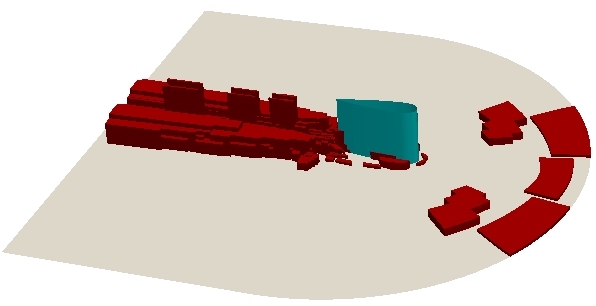}
            \label{fig:c3_res}
            }
          \caption{Case Study III training cases (a) and results (b). The training cases consist of section cuts of the domain (blue) and human-labeled regions containing the horseshoe vortex (red). The results show the entire, mirrored, simulation domain including the wall (grey), airfoil (blue), and the identified vortex (red).}
          \label{fig:c3_io}
        \end{figure}

        A CNN was trained using four section cuts shown in Figure~\ref{fig:c3_train}. 
        The horseshoe vortex was identified visually based on looping streamlines of the cross-flow components of velocity (no mean flow direction) at different cross-sections normal to the airfoil.  
        The CNN architecture consists of inputs size $10\times10\times10$, ten feature maps in the convolutional layer with window size $4\times4\times4$ and max pooling with $2\times2\times1$ windows, and two dense layers of $10$ ReLU and $2$ softmax neurons.  
        Similar to boundary layer case, the training windows were restricted to be the same width as the section cuts. 
        The mini-batches consisted of 5 horseshoe vortex cases, 5 difficult background cases, and 10 background cases with the IoU thresholds being the typical $0.5$ and $0.25$. 
        The region proposal was done with 25 windows of width $3$, height-to-depth ratios of ${1:1,1:2,1:3,1:4,1:5}$ and scales ${10,20,30,40,50}$ cells, with strides of $(4,1,2)$ cells in the $(x,y,z)$ directions, resulting in $181,728$ total regions. 

        The results were evaluated qualitatively and are shown in Figure~\ref{fig:c3_res}. 
        The Fluid R-CNN was able to identify most of the horseshoe vortex, doing particularly well downstream of the airfoil, but not as well around the airfoil where large portions of the vortex were not identified. 
        The identified vortex in the wake also contains erroneous sections near the symmetry plain that extend far from the wall. 
        The method also falsely identifies a region of the wall near the inlet as horseshoe vortex. 
        It is believed that these failures are due to the limited amount of training data, and the simplicity of the CNN architecture used.

\section{Discussions}
    \label{S:discussions}
    The goal of this study was to evaluate the potential of adapting a data-driven algorithm developed for image recognition to the problem of flow feature detection. 
    The results are very encouraging, specially considering the limited amount of training data, the simplicity of the CNN architecture, and all the simplifications done to the overall R-CNN method. 
    There are however some important considerations if this method will continue to be pursued. 

    \subsection{Improving the Fluid R-CNN Method}
        \label{S:disc_improv}
        The Fluid R-CNN method can be improved at every stage using techniques already used in the image recognition counter-parts. 
        The CNN classification can be improved by: (1) using more training examples and artificially augmenting the training data, (2) using deeper architecture and optimizing the architecture and training hyper-parameters, (3) using regularization techniques during training, such as dropout, to avoid over-fitting. 
        For applications requiring identification of several classes of flow features the CNN could be used for local-feature extraction only, with classification done by class-specific support vector machines, allowing for greater specialization of these classifiers. 
        For these cases a more sophisticated region proposal algorithm can be used, which would only pass the most likely regions to the classifiers. 
        For cases with more than one instance of a feature of interest, an IoU threshold can be used on the non-maximum suppression at the selection step to allow for some amount of intersection between selected regions (see Figure~\ref{fig:imgrec}). 
        As can be appreciated from this discussion, the methodology can become very complex very quickly. 
        Although such complexities were not warranted for a first study into the adaptability of these methods to fluid flows, their incorporation should improve the performance significantly. 

        In this study the flows were restricted to flows with singly-connected domains, and were mapped to a rectangular domain. 
        There are several ways of making this a general method. 
        One possible approach would be to divide a multiply-connected domain into several singly-connected domains. 
        The Fluid R-CNN method would first be used on each individually, and then regions along the boundaries which encompass two domains would need to be considered. 
        Another approach we have tried with initial success is to to completely forgo the domain mapping and search the physical domain directly, using rectangular windows of fixed physical dimensions. 
        The biggest challenge of this approach is in handling the boundaries, a problem shared with other fluid dynamics data post-processing techniques such as particle image velocimetry (PIV). 
        Some of the same boundary-handling methods used in PIV could be used for the Fluid R-CNN method. 
        
        Another aspect of the formulation that requires more thought is the identification of continuous flow features, which has no analogue in image recognition. 
        For instance addressing the issue with larger regions (e.g. thicker boundary layer) through either modifying the training data creation step or the selection step. 
        Similarly, the clustering of the identified region into different instances of the flow feature was not considered in this study.

    \subsection{Extension to Time-Dependent Cases}
        As discussed in the introduction, flow feature identification in time-dependent flow fields (e.g. coherent structures in turbulent flows) is an important application and motivation for the development of this method.
        While the current study focused on time-independent problems, the presented methodology can be used directly for identifying flow features at individual time-steps in time-dependent flows.
        Time-dependent datasets have additional problems of interest, including event detection and feature tracking\cite{post_state_2003}. 
        Feature tracking consist of determining if features at different times correspond to the same feature.  
        Events in the time evolution of features include changing shape, splitting into two, dissipating, merging with another feature, and entering or leaving the domain. 
        Two common approaches for feature tracking and event detection include extracting features from the spatial-temporal domain directly (e.g. treating the problem as four dimensional), or by extracting features at each separate time and solving the correspondence problem. 
        The presented method should be useful in either of these approaches.


\section{Conclusions}
    \label{S:conclusions}
    Feature identification is an important task in fluid dynamics applications. 
    Existing methods are feature-specific and are based on physical understanding of the behavior of the flow at regions where the feature occurs. 
    In this study convolutional neural networks, a machine learning technique developed for image recognition, were used to solve the problem of flow feature identification. 
    These techniques work well because the two tasks, object detection in images and flow feature detection, are analogous. 
    In particular, CNNs are well suited for flow feature identification because like objects in images, flow features can occur anywhere within the domain, and CNNs exploit this translation invariance.  
    The novelty of the approach is that it is data-driven, performing the feature detection based on learning from training examples rather than on explicit interpretation of physical laws governing the flow. 
    The method was proven to be successful by applying it to three different case studies: detecting recirculation regions and boundary layer in 2D RANS simulations, and horseshoe vortex in 3D RANS simulations. 
    The results were surprisingly good for the limited amount of training data used and simplicity of the CNN architecture.

    The main advantage of this data-driven method is that it is a general approach to feature extraction rather than being feature-specific, with the ability to detect new features for which a physics based detection method does not exist. 
    Being a data-driven method, it potentially has the ability to distinguish between very similar features, provided enough training data. 
    There are also some challenges with this data-driven method. 
    The method relies on a large amount of human-labeled training data, which can be difficult to obtain. 
    The method is also more computationally intensive than some of the current techniques that rely on calculation of a single quantity at each point. 

    In this study we have demonstrated that data-driven methods can provide a general approach to flow feature extraction. 
    Future work will focus on improving the performance by incorporating the changes discussed in Section~\ref{S:disc_improv} and using the method for identifying flow features in time-dependent flows.

\section*{Acknowledgments}
    The authors thank Dr. Todd Lowe at Virginia Tech for his valuable input throughout the project, Dr. Lian Duan at Missouri S\&T for his support on developing the background material, and Yang Zeng for his help in creating some of the images. 
    This work was in part possible thanks to the Kevin T. Crofton Graduate Fellowship.

\appendix

\section{Input Parameters}
    \label{app:A}
    This appendix summarizes the ten input parameters used from Wang et al.~\cite{wang_physics-informed_2017} and Ling et al.~\cite{ling_evaluation_2015}.
    The definition of the invariants is shown in Table~\ref{tb:inputs}.
    These inputs are non-dimensional and Galilean invariant.
    Each input $q$ is obtained by 

    \begin{equation}
        q = \frac{\hat{q}}{(|\hat{q}| + |q^*|)}
    \end{equation}
    where $\hat{q}$ and $q^*$ are the \emph{raw} input and normalization factor respectively.
    The exception is wall-distance based Reynolds number which does not require a normalization parameter and is given by $q = \hat{q}$.
    The definition of non-orthogonality of velocity and its gradient comes from Gorle et al.~\cite{gorle2012rans}.
    The nomenclature for the variables used in Table~\ref{tb:inputs} is given in Table~\ref{tb:nomenc}.
    Repeated indices imply summation, $D$ denotes the total derivative, $\| \cdot \|$ the matrix norm, and $|\cdot|$ the vector norm.

    \iftoggle{extra_spacing}
        {
            \renewcommand{\baselinestretch}{1.0}
        }{}

    \begin{table}[!ht] 
      \centering
      \caption{Nomenclature for input definition.}
      \label{tb:nomenc}
      \begin{tabular}{c| c}
        Symbol & Definition \\ 
        \hline
        $U_i$ & Mean velocity \\
        $k$ & Turbulent kinetic energy \\
        $u'_i$ & Fluctuation velocity \\
        $\rho$ & Fluid density \\
        $\varepsilon$ & Turbulence dissipation rate \\
        $\mathbf{S}$ & Strain rate tensor \\
        $\boldsymbol{\Omega}$ & Rotation rate tensor \\
        $\nu$ & Fluid viscosity \\
        $d$ &  Wall distance \\
        $\boldsymbol{\Gamma}$ & Unit tangential velocity vector \\
        $L_c$ & Characteristic mean flow length scale\\
        $P$ & Fluid pressure \\
        $x_i$ & Position \\
        $t$ & Time \\
        \hline
      \end{tabular}
    \end{table}

    \begin{table}[!ht] 
      \centering
      \caption{Non-dimensional flow features used as input in the classification.}
        \label{tb:inputs}
        \begin{tabular}{P{5.0cm}| P{3.0cm} | P{4.0cm}}
          Input ($q$) & Raw input ($\hat{q}$) & Normalization factor ($q^*$)  \\ 
          \hline
          Ratio of excess rotation rate to strain rate (Q-criterion) & $\frac{1}{2}(\|\boldsymbol{\Omega}\|^2 - \|\mathbf{S}\|^2)$ & $\|\mathbf{S}\|^2$\\  
          & \\
          Turbulence intensity & $k$ & $\frac{1}{2}U_iU_i$\\ 
          & \\
          Wall-distance based Reynolds number & $\min{\left(\dfrac{\sqrt{k}d}{50\nu}, 2\right)}$ & - \\
          & \\
          Pressure gradient along streamline & $U_k\dfrac{\partial P}{\partial x_k}$ & $\sqrt{\dfrac{\partial P}{\partial x_j} \dfrac{\partial P}{\partial x_j}U_iU_i}$ \\ 
          & \\
          Ratio of turbulent time scale  to mean strain time scale &{ $\dfrac{k}{\varepsilon}$ } & { $\dfrac{1}{\|\mathbf{S}\|} $ } \\ 
          & \\
          Ratio of pressure normal stresses to shear stresses & $\sqrt{\dfrac{\partial P}{\partial x_i}\dfrac{\partial P}{\partial x_i}}$ & $\dfrac{1}{2} \rho\dfrac{\partial U_k^2}{\partial x_k}$\\ 
          & \\
          Non-orthogonality between velocity and its gradient & $\left| U_i U_j\dfrac{\partial U_i}{\partial x_j} \right|$ &{$\sqrt{U_l U_l \, U_i \dfrac{\partial U_i}{\partial x_j}U_k \dfrac{\partial U_k}{\partial x_j}}$ } \\ 
          & \\
          Ratio of convection to production of TKE & $U_i\dfrac{dk}{dx_i}$ & $|\overline{u'_j u'_k} S_{j k}|$\\ 
          & \\
          Ratio of total to normal Reynolds stresses & $\|\overline{u'_iu'_j}\|$ & $k$\\ 
          & \\
          Streamline curvature & {$\left|\frac{D \boldsymbol{\Gamma}}{ |\mathbf{U}| Dt}\right|$} & { $\dfrac{1}{L_c}$ }\\
          \hline                                                        
        \end{tabular}
    \end{table}

    \iftoggle{extra_spacing}
        {
            \renewcommand{\baselinestretch}{\stretchval}
        }{}

\section*{References}
\bibliographystyle{elsarticle-num}
\bibliography{references.bib}

\end{document}